\documentclass[traditabstract]{aa}

\usepackage{graphicx}

\usepackage{txfonts}

\begin{document}

\title{Chemical Evolution of High-Redshift Radio Galaxies
 \thanks{Based on data obtained at the VLT through the ESO programs 076.A-0664 and 077.B-0051}
 }

\author{
		K. Matsuoka
		\inst{1}
		\and
		T. Nagao
		\inst{1,2}
		\and
		R. Maiolino
		\inst{3}
		\and
		A. Marconi
		\inst{4}
		\and
		Y. Taniguchi
		\inst{2}
}

\institute{
		Graduate School of Science and Engineering,
		Ehime University, 2-5 Bunkyo-cho, Matsuyama 790-8577, Japan\\
		\email{
				kenta@cosmos.phys.sci.ehime-u.ac.jp,
				tohru@cosmos.ehime-u.ac.jp
		}
		\and
		Research Center for Space and Cosmic Evolution, Ehime
		University, 2-5 Bunkyo-cho, Matsuyama 790-8577, Japan\\
		\email{
				tani@cosmos.ehime-u.ac.jp
		}
		\and
		INAF -- Osservatorio Astrofisico di Roma, Via di Frascati 33,
		00040 Monte Porzio Catone, Italy\\
		\email{
				maiolino@oa-roma.inaf.it
		}
		\and
		Dipartimento di Astronomia e Scienza dello Spazio, Universita
		degli Studi di Firenze, Largo E. Fermi 2, 50125 Firenze, Italy\\
		\email{
				marconi@arcetri.astro.it
		}
}

\abstract
{
We present new deep optical spectra of 9 high-$z$ radio galaxies (HzRGs)
at $z > 2.7$ obtained with FORS2 on VLT. These rest-frame ultraviolet
spectra are used to infer the metallicity of the narrow-line regions (NLRs) 
in order to investigate the chemical evolution of galaxies in high-$z$ universe.
We focus mainly on the \ion{C}{iv}/\ion{He}{ii} and \ion{C}{iii}]/\ion{C}{iv}
flux ratios that are sensitive to gas metallicity and ionization parameter.
Although the \ion{N}{v} emission has been widely used to infer the gas
metallicity, it is often too weak to be measured accurately for NLRs.
By combining our new spectra with data from the literature, we examine
the possible redshift evolution of the NLR metallicity for 57 HzRGs
at $1 \la z \la 4$. Based on the comparison between the observed emission-line
flux ratios and the results of our photoionization model calculations,
we find no significant metallicity evolution in NLRs of HzRGs, up to $z \sim 4$.
Our results imply that massive galaxies had almost completed their chemical
evolution at much higher redshift ($z > 5$). Finally, although we detect strong
\ion{N}{v} emission lines in 5 HzRGs at $z > 2.7$, we point out that high
\ion{N}{v}/\ion{He}{ii} ratios are not indicative of high metallicities but correspond
to high ionization parameters of gas clouds in NLRs.
}


\maketitle


\section{Introduction}

Chemical evolution of galaxies is one of the most important aspects to
understand the formation and evolution of galaxies, since it is closely
related with the past star formation history of galaxies. The most
straightforward way to investigate the chemical evolution of galaxies is
measuring the metallicity of galaxies at various redshifts and exploring the
systematic trends of the metallicity as a function of redshift (see Maiolino
et al. 2008 and references therein). The gas-phase metallicity of
star-forming galaxies in the local universe can be measured rather easily by
analyzing optical emission line spectra (e.g., [\ion{O}{ii}]$\lambda$3727,
H$\beta$, [\ion{O}{iii}]$\lambda$5007, H$\alpha$, [\ion{N}{ii}]$\lambda$6584;
see Nagao et al. 2006b and references therein). However, this is difficult
at $z > 1$, since these emission lines are
generally very faint and shifted into near-infrared wavelengths where
sensitive spectroscopic observations are more difficult than in the optical.

An alternative approach is to focus on active galactic nuclei (AGNs)
instead of star-forming galaxies. AGNs generally show various emission
lines in rest-frame ultraviolet to infrared wavelengths, that arise in gas clouds
photoionized by the radiation from their central engines. Here we focus on
the rest-frame ultraviolet lines (e.g., \ion{N}{v}$\lambda$1240, \ion{C}{iv}
$\lambda$1549, \ion{He}{ii}$\lambda$1640, \ion{C}{iii}]$\lambda$1909),
since we can easily measure the emission-line
fluxes of high-$z$ AGNs by means of optical spectroscopic observations.
These UV emission lines in AGN spectra are generally brighter than
optical emission lines in star-forming galaxies at similar redshifts.


\begin{table*}[!t]
\caption{Journal of Observations}
\begin{tabular}{l c c c l}

\hline\hline
\noalign{\smallskip}

\multicolumn{1}{c}{Source} &
\multicolumn{1}{c}{$z^a$} &
\multicolumn{1}{c}{$E(B-V)^b$} &
\multicolumn{1}{c}{$\mathrm{Exp.}^c$} &
\multicolumn{1}{c}{Date}\\

\noalign{\smallskip}
\hline
\noalign{\smallskip}

TN J0121$+$1320&		3.516&	0.039&	23400&	2005 Oct 28, Nov 20,\\
&					&		&		&		2006 Sep 17, 20\\
TN J0205$+$2242&		3.506&	0.100&	5400&	2005 Oct 1\\
MRC 0316$-$257&		3.130&	0.014&	10800&	2006 Jul 27, Aug 25, Oct 2\\
USS 0417$-$181&		2.773&	0.048&	2700&	2006 Oct 2\\
TN J0920$-$0712&		2.760&	0.041&	10800&	2006 Apr 3, 4\\
WN J1123$+$3141&	3.217&	0.018&	14400&	2005 Dec 28,\\
&					&		&		&		2006 Apr 5, May 27, Jun 20\\
4C 24.28&			2.879&	0.018&	10800&	2006 Apr 23\\
USS 1545$-$234&		2.755&	0.257&	14400&	2006 Apr 5, 23, 24\\
USS 2202$+$128&		2.706&	0.067&	10800&	2006 Jul 2, Aug 16\\

\noalign{\smallskip}
\hline

\end{tabular}

\begin{list}{}{}
\item[$^a$]
The redshift based on NED (NASA/IPAC Extragalactic Database).
\item[$^b$]  
Galactic extinction [mag] given by Schlegel et al. (1998).
\item[$^c$]
Total exposure time [sec].
\end{list}

\end{table*}

The metallicity of AGNs has been studied extensively, especially by focusing on
the broad-line regions (BLRs). It has been reported that the BLR metallicity
($Z_\mathrm{BLR}$) is typically higher than the solar value
(e.g., Hamann \& Ferland 1992; Dietrich et al. 2003; Nagao et al. 2006c),
reaching as much as $Z_\mathrm{BLR} \sim 15 Z_\mathrm{\odot}$ in extreme
cases (Baldwin et al. 2003b; Bentz et al. 2004). However it is not clear how
the gas metallicity inferred from the broad lines is related to the chemical
properties of the host galaxies, since broad lines of AGNs sample only
very small regions in galactic nuclei ($R_\mathrm{BLR} < 1 \ \mathrm{pc}$; e.g., Suganuma et al. 2006),
which may have evolved more quickly than their host galaxies. For this reason,
here we focus on narrow-line regions (NLRs), instead of BLRs.

Contrary to BLRs, the size of NLRs is roughly comparable to the size of their host galaxies
($R_\mathrm{NLR} \sim 10^{2-4} \ \mathrm{pc}$; e.g., Bennert et al. 2006).
The mass of the NLR is about $M_\mathrm{NLR} \sim 10^{5-7} M_{\odot}$
which is much larger than that of the BLR ($M_{\mathrm{BLR}} \sim 10^2-10^4$
$M_{\odot}$; see, e.g., Baldwin et al. 2003a) and the velocity dispersion of
the NLR emission lines well traces the kinematics of their host galaxies (e.g., Greene \&
Ho 2005; Bennert et al. 2006). The NLR is therefore a good tracer of chemical
properties on the galactic scales.

In order to study NLR metallicities it is necessary to focus on type-2 AGNs,
where BLR and strong continuum emission are obscured and do not affect
the accuracy of narrow line emission measurements.
However only very few optically-selected high-$z$ type-2 QSOs have been
discovered so far. We then focus on NLRs of high-$z$ radio galaxies (HzRGs)
since several of them have been identified even at high-$z$.

Some studies on the metallicity of the NLR in HzRGs have already been carried out in the
past. By studying the emission-line flux ratios of \ion{N}{v}$\lambda$1240/\ion{C}{iv}$\lambda$1549 and
\ion{N}{v}$\lambda$1240/\ion{He}{ii}$\lambda$1640, generally used to measure the metallicity of BLRs
in AGNs, De Breuck et al. (2000) found that the typical metallicity of HzRGs is roughly
$0.4 Z_\mathrm{\odot} < Z_\mathrm{NLR} < 3.0 Z_\mathrm{\odot}$. They also reported a possible metallicity
evolution in their sample: radio galaxies at $z > 3$ have lower metallicity than those at $z < 3$. However,
the emission-line flux of \ion{N}{v} in narrow-line radio galaxies is generally too faint, especially for metal-poor
gas. Therefore, only upper-limits on \ion{N}{v} are available for the majority of the radio galaxies investigated
by De Breuck et al. (2000). Iwamuro et al. (2003) found lower metallicities in the NLRs in HzRGs at
$2.0 < z < 2.6$ than De Breuck et al. (2000), based on their analysis of rest-frame optical emission lines from
their sensitive near-infrared observations. In contrast, Humphrey et al. (2008) recently reported that there is
no significant difference in NLR metallicities between their sample of HzRGs at $z \sim 2.5$ and the lower-$z$
radio galaxies investigated by Robinson et al. (1987). To solve these apparent contradictions, it is mandatory
to carry out further spectroscopic observations of HzRGs.

Nagao et al. (2006a) proposed a new metallicity diagnostic diagram that consists of the \ion{C}{iv}$\lambda$1549,
\ion{He}{ii}$\lambda$1640 and \ion{C}{iii}]$\lambda$1909 emission lines, all of which are moderately strong
in rest-frame ultraviolet spectra of HzRGs, even at low metallicities. They studied the metallicity of NLRs of HzRGs
and reported that the observational data do not show any evidence for a significant evolution of the gas metallicity
in NLRs within the redshift range $1.2 < z < 3.8$, and instead they found a clear trend for more luminous AGNs
to have more metal-rich gas clouds. It should be noted, however, that their sample includes only 5 objects at
$z > 2.7$ (or only 2 objects at $z > 3.0$). Thus, observing more HzRGs is crucial to assess the possible metallicity
evolution.

In this paper we report new spectroscopic observations of 9 HzRGs at $2.7 < z <3.5$. By combining the new data
with the Nagao et al. (2006a) database, we discuss the chemical evolution and the metallicity-luminosity relation
of HzRGs in the $z \sim 1 - 4$ redshift range. We adopt a concordance cosmology with $(\Omega_\mathrm{M},
\Omega_\mathrm{\Lambda}) = (0.3, 0.7)$ and $H_\mathrm{0} = 70 \ \mathrm{km} \ \mathrm{s}^{-1} \ \mathrm{Mpc}^{-1}$.


\section{Observations}

We observed 9 HzRGs at $z > 2.7$ with FORS2 (FOcal Reducer and low dispersion Spectrograph 2)
at the VLT (Very Large Telescope). These targets were selected from the catalog of De Breuck et al. (2000),
excluding HzRGs whose \ion{C}{iv}, \ion{He}{ii} and \ion{C}{iii}] have been already measured.
All the observations were executed in the service mode. The list of the target objects with the observation log
is given in Table 1.

Observations were performed with the 300V dispersion element and the G435 order-sorting filter to cover
the $4450 \ \AA < \lambda_\mathrm{obs} < 8700 \ \AA$ range, which, at the redshift of our targets,
corresponds to the rest-frame ultraviolet and thus includes the UV emission lines needed for abundance determination.
We adopted an on-chip binning mode of $2 \times 2$ with a spectral dispersion of $\approx$ 3.2 \AA $\ \mathrm{pixel}^{-1}$
and a spatial scale along the slit of $\approx$ 0\farcs25 $\mathrm{pixel}^{-1}$. The slit width was 1\farcs0, narrower than
the typical seeing (FWHM $\sim$ 1\farcs4). Individual exposure time was 900 s, and the total integration time for each target
is given in Table 1.

Standard data reduction procedures were performed used available IRAF tasks. Bias was subtracted by using an
averaged bias image and flat-fielding was performed using dome-flat images. Cosmic-ray events were then removed.
The wavelength calibration was done by using sky lines. After sky subtraction, we extracted spectra using apertures
of 2\farcs25 along the slit. The spectra were then flux-calibrated by using the following spectrophotometric standard stars.
Note that the weather condition was good during most of our observations. Examination of the ESO sky monitor database
indicates that most ($>90 \ \%$) of our exposures were taken under photometric conditions. Small variations in the absolute
calibration due to varying seeing and transparency are unimportant given that we bin our data into three broad luminosity bins.


\begin{figure*}[!t]
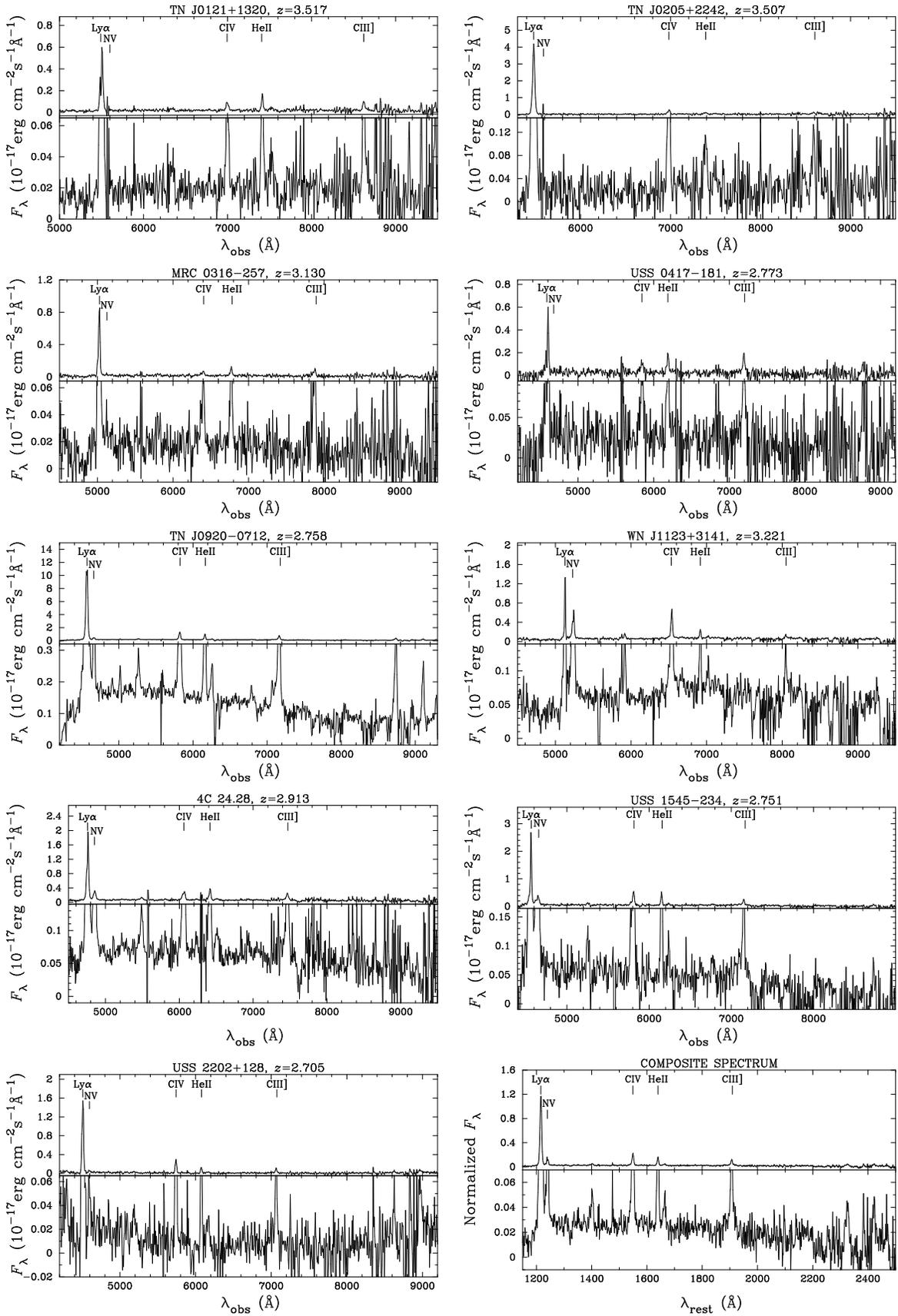

\centering
\begin{tabular}{c c}
\includegraphics[width=7.5cm]{TNJ0121+1320r2.ps}&
\includegraphics[width=7.5cm]{TNJ0205+2242r2.ps}\\
\noalign{\smallskip}
\includegraphics[width=7.5cm]{MRC0316-257r2.ps}&
\includegraphics[width=7.5cm]{USS0417-181r2.ps}\\
\noalign{\smallskip}
\includegraphics[width=7.5cm]{TNJ0920-0712r2.ps}&
\includegraphics[width=7.5cm]{WNJ1123+3141r2.ps}\\
\noalign{\smallskip}
\includegraphics[width=7.5cm]{4C+24.28r2.ps}&
\includegraphics[width=7.5cm]{USS1545-234r2.ps}\\
\noalign{\smallskip}
\includegraphics[width=7.5cm]{USS2202+128r2.ps}&
\includegraphics[width=7.5cm]{Composite2.ps}\\
\end{tabular}
\caption{
Spectra for 9 objects observed with FORS2 and the composite spectrum of the HzRGs
(lower right), improving by binning $\times 2$ in wavelength direction.
Each spectrum is shown on two different flux scales, the first to show strong emission
lines and the second to show continuum emission and weak emission lines.
}
\end{figure*}


\begin{table*}
\caption{Emission-Line Measurements}
\begin{tabular}{l c c c c c}
\hline\hline
\noalign{\smallskip}

\multicolumn{1}{c}{Source}&
\multicolumn{1}{c}{$z^a$}&
\multicolumn{1}{c}{Line}&
\multicolumn{1}{c}{$\lambda_{\mathrm{obs}}$}&
\multicolumn{1}{c}{$\mathrm{Flux}^b$}&
\multicolumn{1}{c}{$\mathrm{FWHM}^c$}\\

&&&
\multicolumn{1}{c}{(\AA)}&
\multicolumn{1}{c}{($10^{-16} \ \mathrm{erg} \ \mathrm{s}^{-1}
\ \mathrm{cm}^{-2}$)}&
\multicolumn{1}{c}{(\AA)}\\

\noalign{\smallskip}
\hline
\noalign{\smallskip}

TN J0121$+$1320&3.517&
\ion{N}{v}$\lambda$1240&--&$<0.094^d$&--\\
&&\ion{C}{iv}$\lambda$1549&6996.9&$0.263\pm0.005$&34.4\\
&&\ion{He}{ii}$\lambda$1640&7414.3&$0.330\pm0.012$&20.1\\
&&\ion{C}{iii}]$\lambda$1909&8622.2&$0.282\pm0.009$&32.6\\

TN J0205$+$2242&3.507&
\ion{N}{v}$\lambda$1240&--&$<0.097^d$&--\\
&&\ion{C}{iv}$\lambda$1549&6981.8&$0.873\pm0.025$&33.0\\
&&\ion{He}{ii}$\lambda$1640&7387.5&$0.519\pm0.046	$&47.3\\
&&\ion{C}{iii}]$\lambda$1909&8586.0&$0.418\pm0.049$&33.4\\

MRC 0316$-$257&3.130&
\ion{N}{v}$\lambda$1240&--&$<0.038^d$&--\\
&&\ion{C}{iv}$\lambda$1549&6396.8&$0.267\pm0.011$&52.4\\
&&\ion{He}{ii}$\lambda$1640&6771.5&$0.301\pm0.009$&27.9\\
&&\ion{C}{iii}]$\lambda$1909&7874.1&$0.345\pm0.018$&41.1\\

USS 0417$-$181&2.773&
\ion{N}{v}$\lambda$1240&--&$<0.103^d$&--\\
&&\ion{C}{iv}$\lambda$1549&5845.1&$0.356\pm0.026$&37.5\\
&&\ion{He}{ii}$\lambda$1640&6188.5&$0.492\pm0.019$&28.9\\
&&\ion{C}{iii}]$\lambda$1909&7192.8&$0.553\pm0.047$&30.7\\

TN J0920$-$0712&2.758&
\ion{N}{v}$\lambda$1240&4663.9&$1.015\pm0.014$&31.6\\
&&\ion{C}{iv}$\lambda$1549&5821.4&$3.365\pm0.010$&26.6\\
&&\ion{He}{ii}$\lambda$1640&6161.5&$2.063\pm0.011$&21.4\\
&&\ion{C}{iii}]$\lambda$1909&7163.2&$1.945\pm0.028$&28.3\\

WN J1123$+$3141&3.221&
\ion{N}{v}$\lambda$1240&5238.7&$1.698\pm0.013$&31.7\\
&&\ion{C}{iv}$\lambda$1549&6538.8&$1.570\pm0.011$&24.8\\
&&\ion{He}{ii}$\lambda$1640&6916.9&$0.425\pm0.014$&19.9\\
&&\ion{C}{iii}]$\lambda$1909&8046.6&$0.183\pm0.028$&20.3\\

4C 24.28&2.913&
\ion{N}{v}$\lambda$1240&4853.2&$1.225\pm0.012$&46.4\\
&&\ion{C}{iv}$\lambda$1549&6061.6&$1.235\pm0.020$&50.7\\
&&\ion{He}{ii}$\lambda$1640&6416.3&$0.978\pm0.011$&29.5\\
&&\ion{C}{iii}]$\lambda$1909&7463.8&$0.812\pm0.041$&41.9\\

USS 1545$-$234&2.751&
\ion{N}{v}$\lambda$1240&4642.9&$1.335\pm0.031$&50.1\\
&&\ion{C}{iv}$\lambda$1549&5810.3&$1.343\pm0.021$&24.7\\
&&\ion{He}{ii}$\lambda$1640&6150.7&$0.878\pm0.012$&16.9\\
&&\ion{C}{iii}]$\lambda$1909&7150.9&$0.606\pm0.031$&26.9\\

USS 2202$+$128&2.705&
\ion{N}{v}$\lambda$1240&4595.0&$0.160\pm0.019$&30.4\\
&&\ion{C}{iv}$\lambda$1549&5739.2&$0.704\pm0.012$&22.9\\
&&\ion{He}{ii}$\lambda$1640&6074.3&$0.289\pm0.010$&20.3\\
&&\ion{C}{iii}]$\lambda$1909&7065.5&$0.292\pm0.011$&25.6\\

\noalign{\smallskip}
\hline

\end{tabular}

\begin{list}{}{}
\item[$^a$]
Redshift of the target calculated from the observed \ion{C}{iv} wavelength.
\item[$^b$]  
Line fluxes corrected for the Galactic reddening.
\item[$^c$]
Observed FWHM (includes instrumental broadening.
\item[$^d$]
3$\sigma$ upper-limit flux.
\end{list}

\end{table*}


\section{Observational results}

The final reduced spectra of the observed HzRGs are shown in Figure 1. In all spectra continuum emission is
very faint; typically S/N$\sim 3$ and ranging $1 \la \mathrm{S/N} \la 10$ at $\lambda_\mathrm{rest} \sim 1700$ \AA.
However the equivalent widths of emission lines are very high and their S/N is accordingly very large.
The composite spectrum of HzRGs, which is shown in the lower right panel of Figure 1, was obtained by stacking
the 9 spectra of HzRGs after converting to rest frame, normalizing to the \ion{He}{ii} flux, and performing a
3 sigma-clipping rejection.

Fluxes of observed emission lines are given in Table 2, where the measurements were performed by fitting
a Gaussian profile with the IRAF task {\tt splot}. The flux error was estimated based on the RMS of the pixels
around each emission line. The detected emission lines are well spectrally resolved and
instrumental broadening is negligible in most cases. Observed emission-line widths without correction
for instrumental broadening are also presented in Table 2. We also provide for each HzRG the redshift calculated
from the \ion{C}{iv} line. For the composite spectrum the flux ratios of \ion{C}{iv}/\ion{He}{ii}, \ion{C}{iii}]/\ion{C}{iv},
and \ion{N}{v}/\ion{C}{iv} are 1.618, 0.628, and 0.584, respectively.

All the emission lines of \ion{C}{iv}, \ion{He}{ii} and \ion{C}{iii}] (required for the metallicity diagnostic method
proposed by Nagao et al. 2006a) are detected in the spectra of all 9 HzRGs observed. In addition, a significant
\ion{N}{v} emission is also detected in 5 out of 9 objects. We provide a 3$\sigma$ upper limit on the \ion{N}{v} emission
for objects without a significant detection, by assuming an emission-line width of $\sim 10 \ \AA$ in rest frame.
As shown in Figure 1, various fainter emission lines are also detected including \ion{Si}{ii}$\lambda$1265,
\ion{O}{i}+\ion{Si}{ii}$\lambda$1305, \ion{C}{ii}$\lambda$1335, \ion{O}{iv}]$\lambda$1402, \ion{O}{iii}]$\lambda$1663,
\ion{Si}{ii}$\lambda$1808, [\ion{Ne}{iv}]$\lambda2422$. We will discuss the properties of these faint emission lines
in a forthcoming paper (Matsuoka et al. in prep.).


\section{Photoionization model}

To infer the metallicity from the observed emission-line spectra, we carried
out model calculations by using the photoionization code Cloudy
version 07.02
\footnote{
We proved for some models that the results of the calculations do not change
significantly with the use of Cloudy version 08.00 instead of version 07.02;
the difference is $\sim 10$ \% at most.
}
(Ferland et al. 1998). Here we assume that the clouds in the NLR of HzRGs
are mainly photoionized and not significantly affected by shocks. Although Nagao
et al. (2006a) demonstrated that this assumption is appropriate when focusing on
\ion{C}{iv}, \ion{He}{ii} and \ion{C}{iii}], we will examine how this assumption
is valid in section 5.1. The parameters for the calculations are (1) the spectral energy
distribution (SED) of the photoionizing continuum radiation; (2) the
hydrogen density of a cloud ($n_\mathrm{H}$); (3) the ionization
parameter ($U$), i.e., the ratio of the ionizing photon density to the
hydrogen density at the irradiated surface of a cloud; (4) the column
density of a cloud ($N_\mathrm{H}$); and (5) the elemental composition
of the gas.


\begin{figure}[!t]
\centering
\includegraphics[width=8cm]{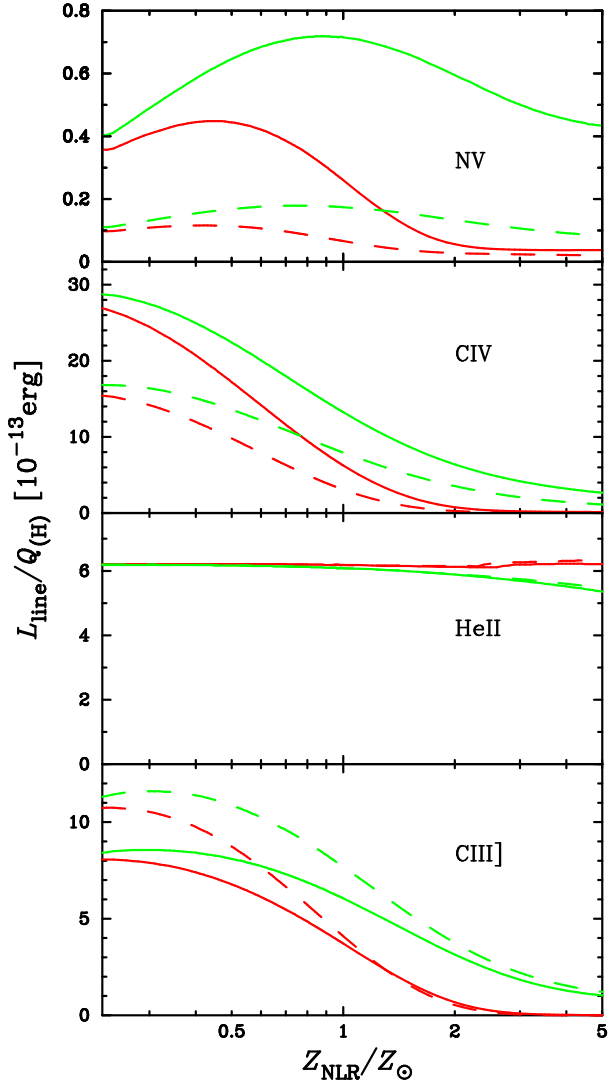}
\caption{
Predicted emission-line luminosity normalized by the number of H-ionizing photons, as functions
of the NLR metallicity. The results for \ion{N}{v}, \ion{C}{iv}, \ion{He}{ii}, and \ion{C}{iii}] are shown,
from top to bottom. Red and Green lines denote the models with hydrogen density of $n_\mathrm{H}
=10^2 \mathrm{cm}^{-3}$ and $10^4 \mathrm{cm}^{-3}$, respectively. Solid and dashed lines denote
the models with ionization parameter of $U = 10^{-1.6}$ and $10^{-2.0}$, respectively.
}
\end{figure}

For the SED of the ionizing photons, we used the "table AGN" command,
that roughly reproduces the typical SED of ionizing photons in AGNs
(Mathews \& Ferland 1987). We do not examine the dependences of the
calculations on the SED but adopt only this SED, because Nagao et al. (2006a)
already investigated the dependences in HzRGs and concluded that the SED effects
do not affect the discussion of the NLR metallicity evolution in HzRGs.
We adopted gas clouds
with the hydrogen density $n_\mathrm{H}=10^2$ and $10^4 \ \mathrm{cm}^{-3}$
and the ionization parameters $U = 10^{-2.4} - 10^{-1.2}$. Here we
assumed dust-free gas clouds, since dusty models are inconsistent
with observations when high-ionization emission lines of HzRGs are concerned
(Nagao et al. 2006a). Note that this is suggested also by rest-frame
optical or near-infrared emission lines (e.g., Marconi et al. 1994;
Ferguson et al. 1997b; Nagao et al. 2003). For the chemical composition
of gas, we assumed that the all metals scale by keeping solar ratios
except for He and N. For helium, we assumed a primary nucleosynthesis
component in addition to the primordial value. Nitrogen scales
as the square power of other metal abundances because
it is a secondary element. We adopted the analytical expressions for the
helium and nitrogen relative abundances as functions of the metallicity
given in Dopita et al. (2000). Another free parameter in our calculation
is the cloud column density. Since we are now focusing on relatively high-ionization
emission lines (\ion{N}{v}, \ion{C}{iv}, \ion{He}{ii}, and \ion{C}{iii}]), we stop our calculations
when the hydrogen ionization fraction drops below 15\%. This requirements ensures that
the line fluxes of interest will not depend on the choice of a particular column density.

In Figure 2 we show the dependence of the emission-line luminosity
on the NLR metallicity, for models with $n_\mathrm{H} = 10^2 \ \mathrm{cm}^{-3}$ and
$10^4 \ \mathrm{cm}^{-3}$, and $U = 10^{-1.6}$ and $10^{-2.0}$. The line luminosity is
normalized by the number of H-ionizing photons in the input
continuum emission. The model behavior is completely different
between \ion{He}{ii} and the other emission lines: the
\ion{He}{ii} is a recombination line and its luminosity is proportional
to the rate of $\mathrm{He}^+$-ionizing photons, while the other emission lines
are collisionally excited and their emissivity
strongly depends on the gas temperature. The equilibrium
temperature of ionized gas clouds is sensitive to
metallicity because of the radiative cooling by metal emission lines.
Therefore the luminosity of collisionally excited emission lines
decreases at high metallicity. Note that the \ion{N}{v} luminosity
decreases at much higher metallicity than the \ion{C}{iv} and \ion{C}{iii}]
luminosity. This is because the nitrogen abundance increases
at high metallicity, i.e., N/H $\propto$ $\mathrm{(O/H)}^2$. All of the above
results are insensitive to the adopted gas density and ionization parameter.

Figure 2 suggests that the \ion{C}{iv}/\ion{He}{ii} flux ratio is possibly good
metallicity diagnostic. However, this flux ratio also depends on
other parameters such as the ionization parameter. 
This degeneracy can be solved by combining it with the \ion{C}{iii}]/\ion{C}{iv} ratio
which is primarily sensitive
to the ionization parameter. Therefore a diagram involving both
the \ion{C}{iv}/\ion{He}{ii} and \ion{C}{iii}]/\ion{C}{iv} ratios is expected to be a powerful
metallicity diagnostic, as originally proposed by Nagao et al. (2006a).
In Figure 3, the results of our photoionization model calculations
are plotted on this diagnostic diagram. The model grids indicate that this diagram is quite
useful to investigate the NLR metallicity. Note that, as shown in
Figure 3, the \ion{C}{iv}/\ion{He}{ii} flux ratio depends also on the gas density
and thus the gas metallicity is not uniquely determined through
this diagram. However, it can be useful to investigate {\it relative}
differences in metallicity and to assess the possible redshift
evolution of the NLR metallicity. In the rest of the paper we only
consider models with $n_\mathrm{H} = 10^4 \mathrm{cm}^{-3}$ (a typical NLR density; see,
e.g., Nagao et al. 2001a, 2002a).


\begin{figure*}
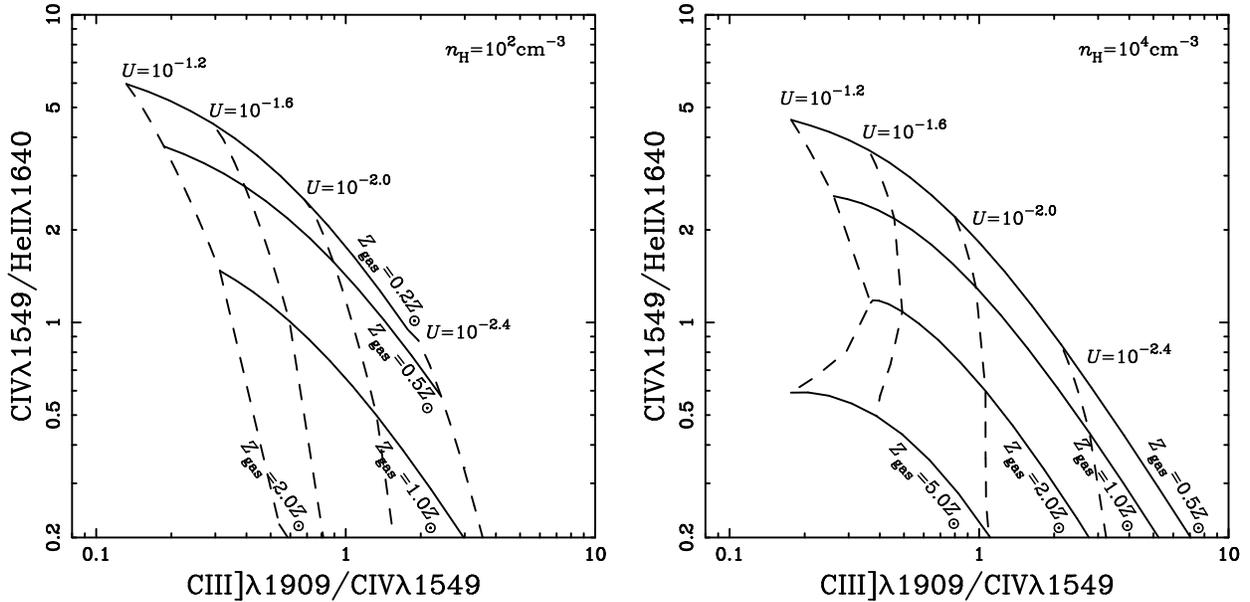

\centering
\begin{tabular}{c c}
\includegraphics[width=8cm]{modelC4He2C3C4n2v3.ps}&
\includegraphics[width=8cm]{modelC4He2C3C4n4v4.ps}\\
\end{tabular}
\caption{
Calculated model grids plotted on a diagram of
\ion{C}{iv}/\ion{He}{ii} versus \ion{C}{iii}]/\ion{C}{iv}. Models for gas
clouds with $n_{\mathrm{H}} = 10^2$ and $10^4 \ \mathrm{cm}^{-3}$ are presented
in the left and right panels, respectively. Constant metallicity and
constant ionization parameter sequences are denoted by solid and dashed lines, respectively.
}
\end{figure*}


\begin{figure*}[!t]
\centering
\includegraphics[width=11.5cm]{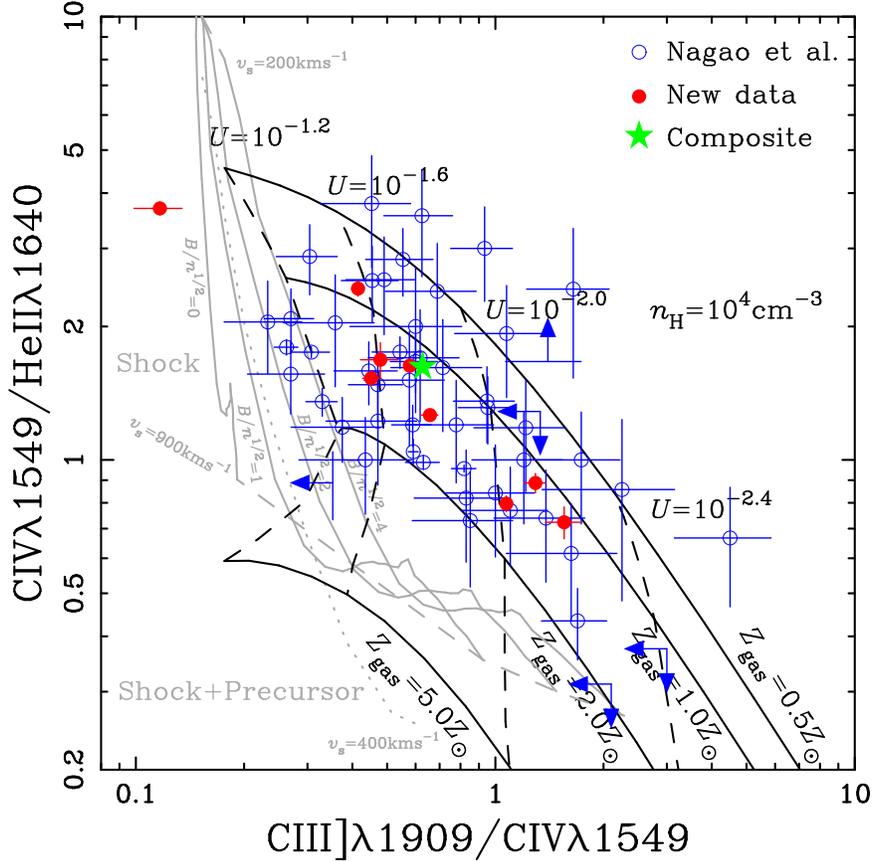}
\caption{
The flux ratios of HzRGs plotted on the diagnostic diagram on \ion{C}{iv}/\ion{He}{ii} versus
\ion{C}{iii}]/\ion{C}{iv}. Our new data are shown with red-filled circles, while the data
compiled by Nagao et al. (2006a) are shown with blue-open circles and arrows.
The green-filled star shows the line ratios measured in the composite spectrum of the 9
HzRGs observed by us. The shock models by Allen et al. (2008) are
also plotted for comparison. Gray-solid lines denote the predictions of pure shock models with magnetic
parameters of $B/n^{1/2}=0$, 1, 2, and 4 $\mu$G $\mathrm{cm}^{3/2}$, solar abundance,
and shock velocities in the range 200 km $\mathrm{s}^{-1} < v_s < 900$ km $\mathrm{s}^{-1}$
(constant velocity sequences of 200 km $\mathrm{s}^{-1}$ and 900 km $\mathrm{s}^{-1}$
are denoted by gray-dashed lines). The gray-dotted line denote the prediction of shock plus precursor models
with $B/n^{1/2}=2$ $\mu$G $\mathrm{cm}^{3/2}$, solar abundance, and 200 km $\mathrm{s}^{-1} < v_s
< 400$ km $\mathrm{s}^{-1}$. Black-solid and black-dashed lines are the same as those in Figure 3.
}
\end{figure*}


\section{Discussion}

\subsection{The metallicity-redshift and metallicity-luminosity relations in HzRGs}

By comparing the observed flux ratios of HzRGs with the prediction of the photoionization models,
we investigate the gas metallicity. In Figure 4, we plot the flux ratios of HzRGs on the diagnostic
diagram with the calculated model grids. We also plot the flux ratios of the composite spectrum of
HzRGs on this diagram. Though new 9 data are at higher redshift than the data in Nagao et al.
(2006a) on average, there is no systematic difference in the data distribution between our new
observations and the data of Nagao et al. (2006a). This result naively suggests that there is no
significant chemical evolution in NLRs of HzRGs, even up to $z \sim 4$.

We now investigate whether shock models can explain the observed flux ratios of HzRGs. In Figure 4 we overplot
the shock models and shock plus precursor models presented by Allen et al. (2008). Specifically,
we plot shock models with shock velocities of 200 km $\mathrm{s}^{-1} < v_s < 900$ km
$\mathrm{s}^{-1}$, magnetic parameters of $B/n^{1/2}=0$, 1, 2, and 4 $\mu$G $\mathrm{cm}^{3/2}$,
and solar abundance, and shock plus precursor models with shock velocities of 200 km
$\mathrm{s}^{-1} < v_s < 400$ km $\mathrm{s}^{-1}$, a magnetic parameter of $B/n^{1/2}=2$ $\mu$G
$\mathrm{cm}^{3/2}$, solar abundance. As shown in Figure 4, both pure shock models and shock plus
precursor models can not explain the observed distribution of line ratios, at variance with pure photoionization models.
We therefore conclude that the dominant ionization mechanism of gas clouds emitting
\ion{C}{iv}, \ion{He}{ii}, and \ion{C}{iii}] is photoionization.

To investigate the relationship between metallicity and redshift or
luminosity, we divided our sample into the following three redshift bins:
$1.2 < z < 2.0$, $2.0 < z < 2.5$, and $2.5 < z < 3.8$.
As for the luminosity, since the non-thermal AGN continuum is obscured in HzRGs,
we used the \ion{He}{ii} line luminosity as a proxy for the AGN's bolometric luminosity.
This directly traces the rate of $\mathrm{He}^+$-ionizing photons which is proportional
to the bolometric luminosity. Luminosity bins are as follows:
$41.5 < \log L$(\ion{He}{ii}) $< 42.5$, $42.5 < \log L$(\ion{He}{ii}) $< 43.0$ and,
$43.0 < \log L$(\ion{He}{ii}) $< 45.0$ (see, Table 3). In Figures 5 and 6, we
investigate the averaged flux ratios of \ion{C}{iv}/\ion{He}{ii} and
\ion{C}{iii}]/\ion{C}{iv} for each luminosity and redshift bin to examine the
possible metallicity dependence on luminosity and redshift. There is no
clear dependence between the flux ratios and redshift as shown in Figure 5,
suggesting that the NLRs metallicity of HzRGs does not show significant
redshift evolution in the range of $1 \la z \la 4$. On the other hand, in Figure
6 we find that the flux ratios are correlated with the luminosity, suggesting
that more luminous HzRGs have more metal-rich gas clouds.
Figure 7 shows the redshift evolution of average metallicity as derived from our sample.
Here we use only the HzRG samples with $\log L$(\ion{He}{ii}) $=42.5-43.0$ to minimize
possible Malmquist bias. Note that the metallicity shown here are derived by assuming
$n_\mathrm{H}=10^4 cm^{-3}$ (see Table 3).
Figure 7 clearly shows that there is no significant decline in the NLR metallicity
at high redshift. In the same way, Figure 8 shows the luminosity dependence of
the averaged NLR metallicity in our sample.
Figure 8 clearly shows the correlation between the AGN luminosity
and the NLR metallicity. These results are completely consistent with the results
of Nagao et al. (2006a) but more reliable since our new sample includes a larger number
of HzRGs at high redshift ($z > 2.7$). Note that the positive correlation
between the AGN luminosity and the gas metallicity of ionized clouds is indicated also for the BLR,
that does not show any redshift evolution (e.g., Hamann \& Ferland 1993;
Nagao et al. 2006c; see also Jiang et al. 2007; Juarez et al. 2009).


\begin{figure}
\centering
\includegraphics[width=8cm]{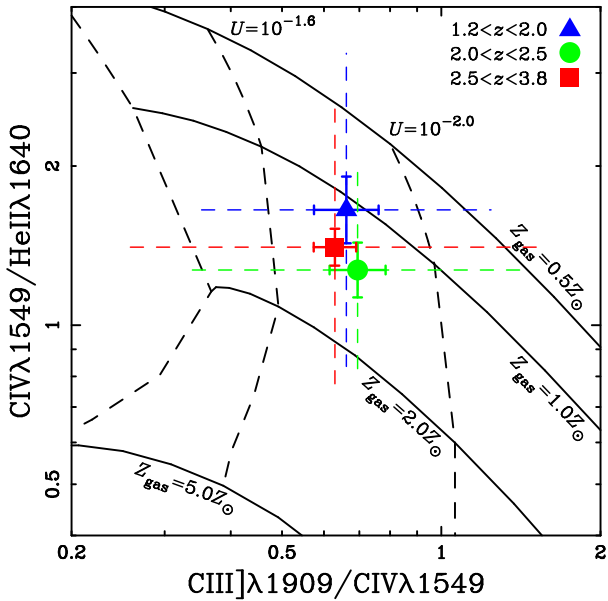}
\caption{
Flux ratios averaged logarithmically for each redshift bin, compared with the model predictions on
the \ion{C}{iv}/\ion{He}{ii} versus \ion{C}{iii}]/\ion{C}{iv} diagram. The dotted bars denote the RMS
of the data distribution, and the solid bars denote the estimated errors on the averaged values.
}
\end{figure}


\begin{figure}
\centering
\includegraphics[width=8cm]{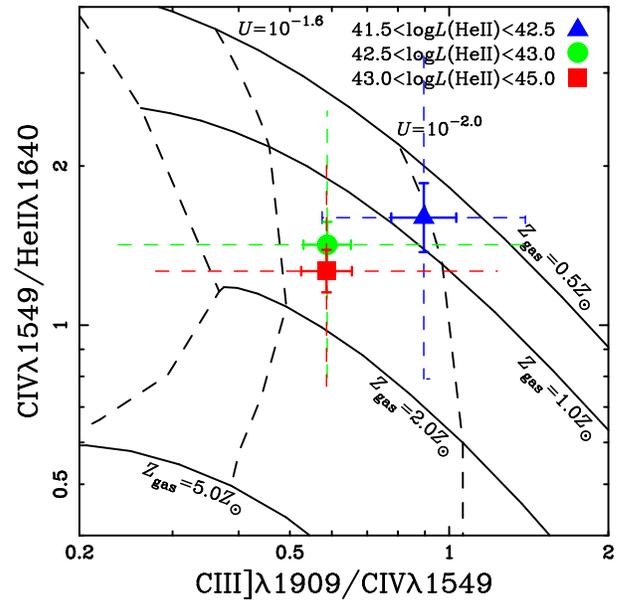}
\caption{
Flux ratios averaged logarithmically for each luminosity bin, compared with the model predictions on
the \ion{C}{iv}/\ion{He}{ii} versus \ion{C}{iii}]/\ion{C}{iv} diagram. The error bars are the same
as in Figure 5.
}
\end{figure}


\begin{table*}
\caption{Averaged Diagnostic Flux Ratios of HzRGs}
\renewcommand{\footnoterule}{}
\begin{tabular}{l l c c c c c c}

\hline\hline
\noalign{\smallskip}

\multicolumn{1}{c}{Sample}&&
\multicolumn{1}{c}{Number}&
\multicolumn{1}{c}{$<z>$}&
\multicolumn{1}{c}{$<\log L($\ion{He}{ii}$)>$}&
\multicolumn{1}{c}{$Z_\mathrm{NLR}/Z_\odot^a$}&
\multicolumn{1}{c}{\ion{C}{iv}/\ion{He}{ii}}&
\multicolumn{1}{c}{\ion{C}{iii}]/\ion{C}{iv}}\\

\noalign{\smallskip}
\hline
\noalign{\smallskip}

$1.2 < z < 2.0$&$41.5 < \log L($\ion{He}{ii}$)^b < 42.5$&8&1.70&41.96&$0.80^{+0.38}_{-0.50}$&$1.82^{+0.47}_{-0.37}$&$0.77^{+0.19}_{-0.16}$\\
\noalign{\smallskip}
&$42.5 < \log L($\ion{He}{ii}$)^b < 43.0$&7&1.66&42.70&$1.27^{+0.34}_{-0.33}$&$1.72^{+0.41}_{-0.33}$&$0.49^{+0.11}_{-0.09}$\\
\noalign{\smallskip}
&$43.0 < \log L($\ion{He}{ii}$)^b < 45.0$&2&1.81&43.32&$1.36^{+0.67}_{-0.60}$&$0.98^{+0.38}_{-0.27}$&$0.99^{+0.40}_{-0.28}$\\
\noalign{\smallskip}
$2.0 < z < 2.5$&$41.5 < \log L($\ion{He}{ii}$)^b < 42.5$&4&2.28&42.25&$0.77^{+0.61}_{-0.77}$&$1.34^{+0.57}_{-0.40}$&$1.14^{+0.43}_{-0.31}$\\
\noalign{\smallskip}
&$42.5 < \log L($\ion{He}{ii}$)^b < 43.0$&10&2.28&42.78&$1.54^{+0.34}_{-0.31}$&$1.21^{+0.22}_{-0.19}$&$0.65^{+0.13}_{-0.11}$\\
\noalign{\smallskip}
&$43.0 < \log L($\ion{He}{ii}$)^b < 45.0$&6&2.24&43.22&$1.53^{+0.28}_{-0.26}$&$1.33^{+0.22}_{-0.19}$&$0.56^{+0.10}_{-0.08}$\\
\noalign{\smallskip}
$2.5 < z < 3.8$&$41.5 < \log L($\ion{He}{ii}$)^b < 42.5$&4&2.78&42.39&$0.88^{+0.25}_{-0.26}$&$1.47^{+0.24}_{-0.21}$&$0.94^{+0.14}_{-0.12}$\\
\noalign{\smallskip}
&$42.5 < \log L($\ion{He}{ii}$)^b < 43.0$&9&3.06&42.73&$1.32^{+0.25}_{-0.22}$&$1.46^{+0.21}_{-0.19}$&$0.61^{+0.10}_{-0.08}$\\
\noalign{\smallskip}
&$43.0 < \log L($\ion{He}{ii}$)^b < 45.0$&7&2.67&43.40&$1.64^{+0.23}_{-0.24}$&$1.30^{+0.16}_{-0.15}$&$0.52^{+0.09}_{-0.08}$\\
\noalign{\smallskip}

\noalign{\smallskip}
\hline

\end{tabular}

\begin{list}{}{}
\item[$^a$]
Assuming the hydrogen density of $n_\mathrm{H}=10^4 \mathrm{cm}^{-3}$.
\item[$^b$]
Line luminosity of \ion{He}{ii}$\lambda$1640 in units of ergs $\mathrm{s}^{-1}$.
\end{list}

\end{table*}


\begin{figure}
\centering
\includegraphics[width=8cm]{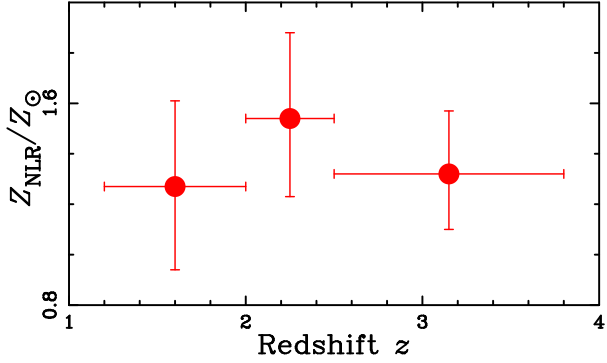}
\caption{
The NLR metallicity versus redshift diagram, plotting metallicities of three subsamples
with $\log L$(\ion{He}{ii}) $= 42.5-43.0$ at
redshift range as follows: $1.2 < z < 2.0$, $2.0 < z < 2.5$, and $2.5 < z < 3.8$ (see also Table 3).
These metallicities and the errors are estimated from the \ion{C}{iv}/\ion{He}{ii} versus
\ion{C}{iii}]/\ion{C}{iv} diagram. Transverse bars show redshift ranges.
}
\end{figure}


\begin{figure}
\centering
\includegraphics[width=8cm]{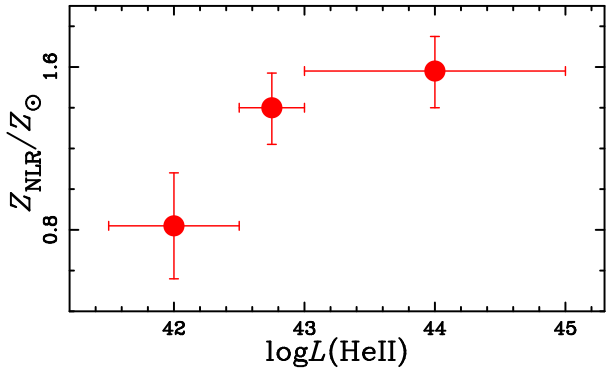}
\caption{
The NLR metallicity versus luminosity diagram, plotting metallicities of three subsamples
at luminosity range as follows: $41.5 < \log L$(\ion{He}{ii}) $< 42.5$, $42.5 < \log L$(\ion{He}{ii})
$< 43.0$ and, $43.0 < \log L$(\ion{He}{ii}) $< 45.0$. These metallicities and the errors are
estimated from the \ion{C}{iv}/\ion{He}{ii} versus \ion{C}{iii}]/\ion{C}{iv} diagram. Transverse bars
show luminosity ranges.
}
\end{figure}

\subsection{Interpretation}

In section 5.1, we reported a significant positive correlation between
the NLR metallicity and the AGN luminosity. In this section we discuss two
possible origins for this correlation:
\begin{itemize}
\item
the galaxy mass-metallicity relation (assuming $L_\mathrm{AGN} \propto
M_\mathrm{BH} \propto M_\mathrm{host}$)
\item
a dependence of NLR metallicity on the Eddington ratio
\end{itemize}
We discuss each of these scenarios below.

One possible scenario for the positive correlation between
NLR metallicity and AGN luminosity is that it
reflects the relation between galaxy mass and metallicity
(e.g., Lequeux et al. 1979; Tremonti et al. 2004; Lee et al. 2006).
Since the metal content of NLRs is the result of the past star-formation
history in the host galaxies, the metallicity of galaxies and that of
NLRs should be closely related. Galaxies are characterized by a well defined mass-metallicity
relation even at high redshift (Maiolino et al. 2008). On the other hand, AGN
luminosities and host galaxy masses should also be closely
related, if the Eddington ratio is roughly the same within this class of objects and the
correlation between the mass of
supermassive black holes (SMBHs) and that of galaxy spheroidal
components holds also at high redshift. Therefore, by assuming a
narrow range of the Eddington ratios and the $M_\mathrm{BH}$-galaxy mass relation,
the positive correlation between NLR metallicity and AGN
luminosity is naturally expected. Indeed, some observational
studies report that the Eddington ratio of high-z quasars is limited
in a narrow range (e.g., Kollmeier et al. 2006; Trump et al. 2009).

However, this scenario has a serious problem. We found that the
NLR metallicity shows no significant redshift evolution, up to $z\sim4$.
On the contrary, the mass-metallicity relation in galaxies shows
a significant redshift evolution, at least in the $0 < z < 3$ range
(e.g., Savaglio et al. 2005; Erb et al. 2006; Liu et al. 2008; Maiolino et al. 2008).
If the relation between NLR metallicity and AGN luminosity is simply caused
by the galaxy mass-metallicity relation, the NLR metallicity is also expected to show
a significant redshift evolution, in contrast with our observational results.
Below we discuss possible reasons for this apparent contradiction and the implications of the relation
between NLR metallicity and AGN luminosity.

A possibility to reconcile this problem is that the apparent
lack of evolution is the extreme consequence of galaxy downsizing
on metal enrichment. We know that low mass galaxies
evolve slowly, even from the chemical point of view, on a prolonged time scale
extending to the current epoch. In contrast, massive galaxies reach their chemical
maturity on short time scales and at high redshift (Maiolino et al. 2008
and references therein). Since the host galaxies of HzRGs are very massive
(e.g. De Breuck et al. 2002), their chemical evolution may be completed
at much earlier epochs than observed in the current sample, i.e. at $z > 3$.
This scenario predicts that the NLR metallicities in lower-luminosity HzRGs
(hence probably hosted in less massive hosts) may show evolution even
at $z < 3$. This prediction can be tested with sensitive spectroscopic observations
of faint HzRGs.

Alternatively, the relation between NLR metallicity and AGN luminosity may be
independent of the mass-metallicity relation of galaxies, since the Eddington ratio
is likely not universal.
Recently it was reported that the BLR metallicity in quasars is correlated with
the Eddington ratio (e.g., Shemmer et al.
2004; see also Nagao et al. 2002b; Shemmer \& Netzer 2002).
This suggests that the NLR metallicity may be also correlated with the Eddington
ratio, although the physical origin is not clear. In this case, the correlation between
the AGN luminosity and the NLR metallicity is independent of the galaxy mass-metallcity
relation.

Whatever the scenario, our study strongly suggests that HzRGs completed
their major chemical evolution in the very high-z universe, i.e., $z > 4$.
If the minimum time-scale for a significant enrichment of carbon
($\sim0.5$ Gyr; e.g., Matteucci 2008) is taken into account, the major
epoch of the star formation in HzRGs may have occurred at $z > 5$.

Note that our conclusions are in contrast with some of the earlier studies
mentioned in section 1. The most significant
difference between previous works and our own is
the adopted diagnostics, i.e., we do {\it not} use \ion{N}{v} emission to investigate
the NLR metallicity. While the \ion{N}{v} emission is too weak in most HzRGs
to be measured accurately, all of the lines used by us (\ion{C}{iv}, \ion{He}{ii},
and \ion{C}{iii}]) are so strong that we can measure their fluxes rather easily.
Therefore we can investigate the chemical evolution of HzRGs without
relying on upper-limit data. Another serious problem in using the \ion{N}{v}
emission in metallicity studies is that the flux ratios including
\ion{N}{v} are sensitive not only to the NLR metallicity but also to the ionization parameter
as discussed in the following section.


\begin{figure}
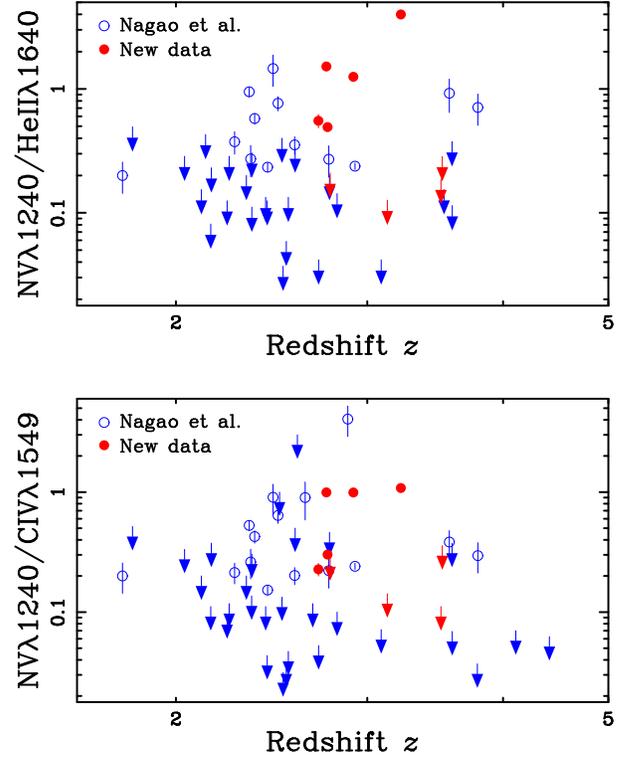

\centering
\begin{tabular}{c}
\includegraphics[width=8cm]{FIG-RLT-n4-N5He2z.ps}\\
\noalign{\bigskip}
\includegraphics[width=8cm]{FIG-RLT-n4-N5C4z.ps}\\
\end{tabular}
\caption{
The flux ratios of new sample plotted on diagram of \ion{N}{v}/\ion{He}{ii} (upper)
and \ion{N}{v}/\ion{C}{iv} (lower). Our new data are shown with the red-filled
circles, and the previous data compiled by Nagao et al. (2006a) are shown
with the blue-open circles and arrows.
}
\end{figure}

\subsection{The \ion{N}{v} emission in HzRGs}

Given the high S/N of our spectra, the \ion{N}{v} emission line is securely detected
in 5 out of 9 objects. Nitrogen is a secondary element and thus its abundance is
expected to be proportional to the square of the metallicity [i.e., N/O $\propto$ O/H, or
equivalently, N/H $\propto$ $\mathrm{(O/H)}^2$].
Therefore, the \ion{N}{v} line is widely regarded as a good metallicity indicator, at least
for BLRs (e.g., Hamann \& Ferland 1992, 1993; Hamann et al. 2002; Dietrich et al. 2002,
2003; Nagao et al. 2006c). Accordingly the \ion{N}{v} line is sometimes regarded as
a metallicity indicator also for NLRs (e.g., van Ojik et al. 1994; De Breuck et al. 2000;
Vernet et al. 2001; Overzier et al. 2001; Humphrey et al. 2008). We thus examine
the \ion{N}{v}/\ion{C}{iv} and \ion{N}{v}/\ion{He}{ii} flux ratios as a function of redshift, to
infer possible redshift evolution of the NLR metallicity independently of the diagnostic
diagram of \ion{C}{iv}/\ion{He}{ii} versus \ion{C}{iii}]/\ion{C}{iv}. These two flux ratios are
plotted as a function of redshift in Figure 9. Our new data at $z > 2.7$ clearly show
the existence of HzRGs with high \ion{N}{v}/\ion{C}{iv} and \ion{N}{v}/\ion{He}{ii} flux ratios
even at $z > 2.7$, which seems consistent to the idea that there is no significant evolution
in the NLR metallicity of HzRGs as suggested by the \ion{C}{iv}/\ion{He}{ii} and
\ion{C}{iii}]/\ion{C}{iv} flux ratios.

Figure 9 shows, however, the presence of a large number of HzRGs with a very weak
\ion{N}{v} emission. In particular, some HzRGs show \ion{N}{v}/\ion{He}{ii} $< 0.1$ and
\ion{N}{v}/\ion{C}{iv} $< 0.1$ while others show \ion{N}{v}/\ion{He}{ii}
$\sim 1$ and \ion{N}{v}/\ion{C}{iv} $\sim 1$. With the models presented by
Vernet et al. (2001), these flux ratios would correspond to $Z_\mathrm{NLR} < 0.4 Z_\mathrm{\odot}$
for \ion{N}{v}-weak HzRGs and $Z_\mathrm{NLR} \sim 4 Z_\mathrm{\odot}$
for \ion{N}{v}-strong HzRGs. This would suggest that the NLR metallicity of HzRGs varies from
object to object in a very wide range ($> 1$ dex). This seems inconsistent with
the results obtained through the analysis of the \ion{C}{iv}/\ion{He}{ii} and
\ion{C}{iii}]/\ion{C}{iv} emission-line ratios (Figure 3). To investigate the origin of this
apparent contradiction, in Figure 10 we plot the observational data on the \ion{N}{v}/\ion{He}{ii}
versus \ion{N}{v}/\ion{C}{iv} diagram and compare them with our model calculations.
The distribution of the observational data on this diagram seems consistent with the area spanned by
model grids. The near horizontal lines of constant-$U$ in Figure 10 suggest that
the \ion{N}{v}/\ion{He}{ii} flux ratio is sensitive to the ionization parameter but insensitive to
the NLR metallicity. The \ion{N}{v}/\ion{C}{iv} ratio appears to depend on both
$U$ and $Z_\mathrm{NLR}$. In other words, the increase of the ionization
parameter for a given metallicity results in high \ion{N}{v}/\ion{He}{ii} and \ion{N}{v}/\ion{C}{iv}
flux ratios, while the increase of the metallicity for a given ionization parameter results in high
\ion{N}{v}/\ion{C}{iv} but nearly constant \ion{N}{v}/\ion{He}{ii} flux ratios.
This suggests that the \ion{N}{v}/\ion{He}{ii} flux ratio is sensitive to the ionization parameter
rather than to the NLR metallicity. This is demonstrated more clearly in Figure 11,
where we show the \ion{N}{v}/\ion{He}{ii} versus \ion{C}{iii}]/\ion{C}{iv} diagram. The narrow coverage
of the model grids in this diagram suggests that both \ion{N}{v}/\ion{He}{ii} and
\ion{C}{iii}]/\ion{C}{iv} flux ratios depends mainly on the ionization parameter but
are nearly independent of the NLR metallicity. Note that the observational data appears to
be distributed in a wider range with respect to the one predicted by the models.
Indeed, only the models with $n_\mathrm{H}=10^4 \ \mathrm{cm}^{-3}$ are
plotted in the diagram and most of the observational data are upper limits.


\begin{figure}
\centering
\includegraphics[width=8cm]{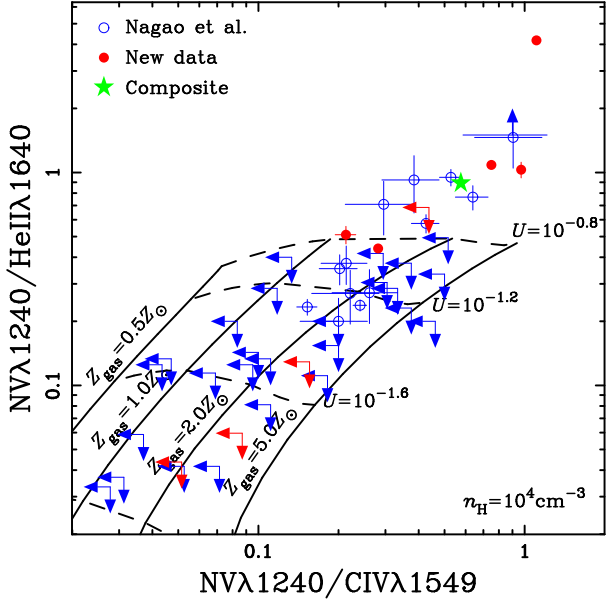}
\caption{
The flux ratios of HzRGs plotted on the diagnostic diagram on \ion{N}{v}/\ion{He}{ii} versus
\ion{N}{v}/\ion{C}{iv}. Red-filled circles, blue-open circles, arrows and green-filled star are
the same as those in Figure 3. Solid and dashed lines are the same as those in Figure 2.
}
\end{figure}


\begin{figure}
\centering
\includegraphics[width=8cm]{resultN5He2C4C3n4v2.ps}
\caption{
The flux ratios of HzRGs plotted on the diagnostic diagram on \ion{N}{v}/\ion{He}{ii} versus
\ion{C}{iii}]/\ion{C}{iv}. Red-filled circles, blue-open circles, arrows and green-filled star are
the same as those in Figure 3. Solid and dashed lines are the same as those in Figure 2.
}
\end{figure}


\begin{figure}
\centering
\includegraphics[width=8cm]{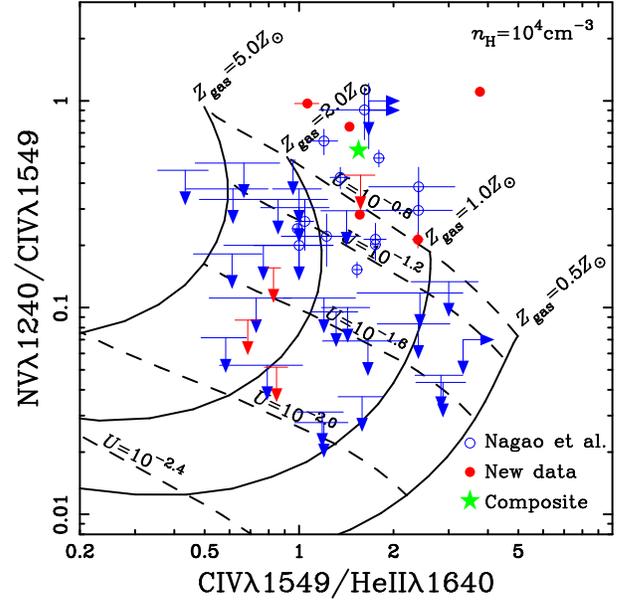}
\caption{
The flux ratios of HzRGs plotted on the diagnostic diagram on \ion{N}{v}/\ion{C}{iv} versus
\ion{C}{iv}/\ion{He}{ii}. Red-filled circles, blue-open circles, arrows and green-filled star are
the same as those in Figure 3. Solid and dashed lines are the same as those in Figure 2.
}
\end{figure}

The physical reason why the \ion{N}{v}/\ion{He}{ii} flux ratio is less sensitive
than the \ion{N}{v}/\ion{C}{iv} flux ratio is shown in Figure 2 and described in section 4.
Both \ion{N}{v} and \ion{C}{iv} are collisionally excited emission lines and thus become
weak at high metallicity where the equilibrium temperature is low.
However, \ion{N}{v} emission decreases much more slowly with metallicity
than \ion{C}{iv}, because nitrogen is a secondary element.
This is the main reason of the metallicity dependence of the \ion{N}{v}/\ion{C}{iv} flux ratio.
However, especially at $U > 10^{-2.0}$, the metallicity dependence of
the \ion{N}{v} emission is small (see the top panel of Figure 2).
As already mentioned in section 4, the \ion{He}{ii} emission is also
insensitive to the metallicity because it is a recombination line.
This is demonstrated in Figure 12, where we plot the \ion{N}{v}/\ion{C}{iv} versus
\ion{C}{iv}/\ion{He}{ii} diagram. Here $Z_\mathrm{NLR}$-sequences are mostly
perpendicular to $U$-sequences, and $Z_\mathrm{NLR}$-sequences go towards the
upper-left direction with increasing NLR metallicity. However
this diagram is less useful than the \ion{C}{iv}/\ion{He}{ii} versus \ion{C}{iii}]/\ion{C}{iv}
diagram to infer the NLR metallicity, because \ion{N}{v}/\ion{C}{iv} flux ratios of
HzRGs are difficult to be measured accurately; indeed most of the observational
data in Figure 12 are upper-limit values.

In Figure 9 -- 12, the inferred ionization parameter for the \ion{N}{v}-detected HzRGs is much
higher than that suggested by the diagram in Figure 4. This is due to the higher ionization
potential of the $\mathrm{N}^{4+}$ ion (77.4 eV) compared to the $\mathrm{C}^{3+}$ (47.9 eV)
and $\mathrm{He}^{2+}$ (54.4 eV) ions; i.e., the \ion{N}{v} emission tends to arise from
regions characterized by higher ionization parameters with respect to \ion{C}{iv} and
\ion{He}{ii}. Although multi-zone photoionization models for NLRs (e.g., Ferguson
et al. 1997a; Nagao et al. 2001b) solve this discrepancy in ionization parameter,
it is beyond the scope of this paper and thus they are not discussed further.


\section{Conclusion}

To investigate the metallicity evolution of galaxies in the high-$z$ universe,
we have performed new spectroscopic observations of 9 HzRGs at $z > 2.7$, which
complement data available at lower redshifts. By comparing the total set of data of 57 HzRGs at $1 \la z \la 4$
with photoionization models, we found
the following results.

\begin{enumerate}

\item
Our analysis of the emission-line flux ratios of \ion{C}{iv}, \ion{He}{ii} and
\ion{C}{iii}] suggests that there is no significant chemical evolution in the
redshift range of $1 \la z \la 4$.

\item
We found a positive correlation between the NLR metallicity and
the AGN luminosity. There are two possible origins for this correlation:
the galaxy mass-metallicity relation and a dependence of the NLR metallicity
on the Eddington ratio.

\item
The non-evolution of the gas metallicity of the NLRs implies that the major
epoch of star formation in the host galaxies is at $z > 5$.

\item
We detected the \ion{N}{v} emission in the spectra of 5 HzRGs at
$z \ga 2.7$, and found that there are some HzRGs with high \ion{N}{v}/\ion{C}{iv}
and \ion{N}{v}/\ion{He}{ii} flux ratios even at such high redshift.

\item
However, high \ion{N}{v}/\ion{He}{ii} ratios do not always mean high NLR metallicities,
but correspond mainly to high ionization parameters.

\end{enumerate}


\begin{acknowledgements}

We would like to thank the VLT staff for their invaluable help during the observations
and G. Ferland for providing the excellent photoionization code Cloudy to the public.
We also thank the anonymous referee for useful comments and suggestions.
This work was financially supported in part by the Japan Society for the Promotion of Science
(Nos. 15340059, 17253001, and 19340046).
KM acknowledges financial support from the Circle for the Promotion
of Science and Engineering of Japan. TN is financially supported through the Research
Promotion Award of Ehime University. RM and AM acknowledge partial support
by the Italian Space Agency (ASI) through contract ASI-INAF I/016/07/0.

\end{acknowledgements}


\end{document}